# AN INVERSE STEFAN PROBLEM RELEVANT TO BOILOVER:
## Heat Balance Integral Solutions and Analysis

by

### *Jordan HRISTOV*

*Stefan problems relevant to burning oil-water systems are formulated. Two moving boundary sub-problems are defined: burning liquid surface and formation of a distillation ("hot zone") layer beneath it. The basic model considers a heat transfer equation with internal neat generation due to radiation flux absorbed in the fuel depth. Inverse Stefan problem corresponding to the first case solved by the heat balance integral method and dimensionless scaling of semi-analytical solutions are at issue.*

Key words: *fire, boilover, Stefan problems, heat balance integral, traveling wave-like solution, Koseki's thermal wave, critical fuel depth, time to boilover*

## Introduction

The "boilover" occurs when the burning fuel is expelled violently from the tanks due to the vaporization of the underlying water, usually collected due to condensation effects [1, 2]. Number of systematic studies have addressed the fuel layer-waterbed parameters [3-7] predefining the boilover appearance and intensity. The term, commonly referred as "thin-layer boilover", has been applied also to the burning of thin slicks of oil spills spread after accidental releases [5, 7] – see fig.1.

Boilover problem have been studied intensively in experiments [8-14] and by models [1, 12-17, 18 (present issue)] towards deep understanding of the controlling phenomena. Experimental results developed before 1993 were reviewed by Koseki [19], while scale of analysis of models was performed recently in [20, 21]. The heating models of fuel-water stratified layers are based on a general assumption of heat conduction [1, 9, 14-16, 22] and ignored fuel convection. The present work address heat conduction equations defined in [21] as *in-depth absorption model* (DAM) with a volumetric heating sources and *surface absorption model* (SAM) in cases with both linear and non-linear boundary conditions (with a phase-change at the flaming surface).

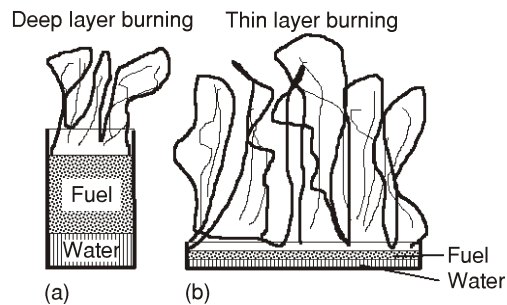

**Figure 1**

The Stefan boundary conditions at the flaming surface [21] and volumetric heating (DAM) make the problem complex and needs approximate or numerical methods [7]. The present work especially applies analytical methods allowing scaling, generation of dimensionless group, and investigation of asymptotic behaviour of the solutions. An *Inverse Stefan Problem* [24] is solved here by the *heat balance integral* (HBI) method [22-24] with a front-fixing technique (FFT) in *ablation approximation* [23] in absence of hot zone beneath the surface. Special attention is paid on the correct interpretation of the equations and mathematical artifacts [23-25] resulting from the FFT approach.

The burning liquid layer (see fig. 2) (decreases in depth) with a regression rate $r(t)$:

$$V_a \quad \dot{y}(t) \quad \frac{\partial y_s(t)}{\partial t} \quad \frac{\dot{m}}{\rho_F} \tag{1}$$

that is a function of the fuel properties and the vessel diameter [5-7, 26, 27] (see also [18]) but it is independent of the fuel layer thickness. The net heat feedback from the flame (per unit area) to the surface of the burning liquid is a fraction $\chi$ of the total heat released [27, 28]:

$$\dot{q}_s \quad \frac{4\chi}{\pi} \rho_\infty C_p \sqrt{T_\infty^f g (T_f \quad T_\infty^f)} \sqrt{D} \tag{2}$$

where $\chi$ $[(1 - e^{-KD})/D^{1/2}]^{0.61}$ [26] and $K$ is extinction coefficient.

Hereafter, we will assume that $\dot{q}_s$ is constant for a given fuel and pool diameter and for seek of simplicity it is denoted as $\dot{q}_s$ $F$ too. Referring to fig. 2a we may define two moving boundary problems relevant to the fuel-water layer heating, namely:

– *inverse 1-D Stefan problem* considers the fuel-water heating history with known and sharp phase-change line (fuel surface) (fig. 2b) defined as *ablation approximation* [23, 24]. The correct term is *inverse Stefan problem* [25] since it addresses the temperature profile across the fuel layer. In the opposite case when the velocity should be defined we have to solve the *forward (direct) Stefan problem* [25]. This simplified situation corresponds to combustion of fuel that do not form isothermal layer beneath the surface (hot zone).

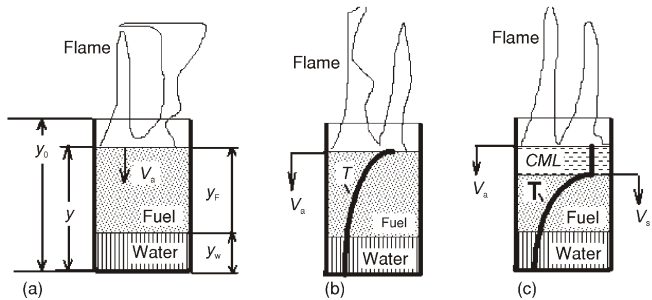

**Figure 2**

– *compound Stefan problem* with a second moving boundary due to the formation of a *hot zone* (HZ) (or *convection mixed layer* (CML), or *isothermal layer, distillation zone*) beneath the flaming surface (see fig. 2c) .The hot zone depth increases in time with a velocity that really defines a *direct Stefan problem* since both the front line of CML and the temperature profile are unknown. This compound problem is beyond the subject of the present discussion.

Hristov *et al.* [20] reviewed the existing 1-D heat transfer equation and performed large scaling studies. The present work addresses the same problem with the HBI and FFT techniques. The one-dimensional in-depth absorption model, ignoring fuel convection, discussed in [20], is:

$$\frac{\partial T}{\partial t} = a_F \frac{\partial^2 T}{\partial y^2} - \frac{\partial \dot{q}_r}{\partial y}, \quad \dot{q}_r = \dot{q}_s e^{-\mu y} \tag{3a}$$

with

$$t = 0, \quad T = T_\infty \quad \text{and} \quad T = T_s, \quad y = y_s(t) \tag{3b}$$

and Stefan boundary condition (SBC) at the liquid surface:

$$\dot{q}_s = H_V \rho_F r(t) + \dot{q}_c, \quad q_c = -\lambda \left. \frac{\partial T}{\partial y} \right|_{y = y_s} \tag{4}$$

As a first approximation we assume [5-7] no differences in thermal properties of fuel and water. The water heating poses a serious question about the limit of superheat since the experimental measurements [5, 18, 20] and the physical experiments [28, 29] provide different data.

## Inverse Stefan problem and front-fixing techniques (FFT)

A moving boundary problem can be easy transformed into a fixed-boundary one by a simple motion of the interface with attached reference frame (see fig. 3a) towards the flame at the velocity [25, 30]. For instance, the FFT contribute the SAM equations with a term $(V_a \rho_F C_{pF}) dT/dy$ [30]:

$$\rho_F C_{pF} \frac{\partial T}{\partial t} + V_a \rho_F C_{pF} \frac{dT}{dy} = \lambda_F \frac{\partial^2 T}{\partial y^2} \tag{5}$$

with $y = 0$, $T = T_s$, and $x = \infty$, $T = T_\infty$, and $dT/dy = 0$       (6)

For constant properties $\lambda_F$, $\rho_F$, and $C_{pF}$ eq. (5) has a closed form solution [30], namely:

$$\frac{T - T_\infty}{T_s - T_\infty} = e^{-\frac{V_a y}{a_F}} \tag{7}$$

Similarly the *in-depth absorption model* (eq. 3a) with FFT becomes:

$$\frac{\partial T}{\partial t} \quad V_a \frac{\partial T}{\partial y} \quad a_F \frac{\partial^2 T}{\partial y^2} \quad \frac{1}{\rho_F C_{pF}} \frac{\partial \dot{q}_r}{\partial y} \qquad (8)$$

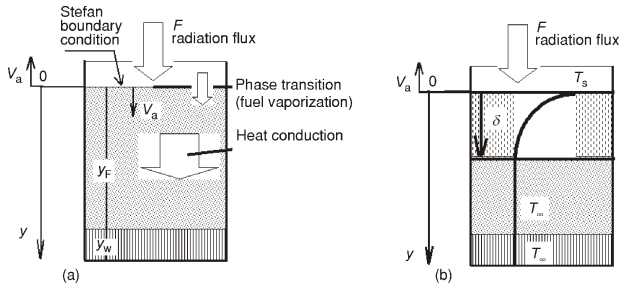

**Figure 3**

The added enthalpy flux term ($V_a \rho_F C_{pF}$) $T$ $y$ in eq. (8) is associated with the negative $V_a$ (opposite to the orientation of the $y$-axis) and is a result of the change of the reference frame. It cannot be associated with any convective motions in the liquids as it was done in [1, 17]. More to the point, $V_a$ can contribute to dimensionless number $N_{DHS} = y_0 r(t)/a_F = y_0 V_a/a_F$ [20] defined only by the boundary conditions but never to the Reynolds number that is dependent on the bulk performance of the liquid). The model (eqs. 3-4) considers *head conduction as a transport mechanism only*. The physical derivation of SBC (eq. 4) considers conservation of energy in which the two phases are assumed to be incompressible and *at rest* [25]. With two phases of different densities $\rho_1$ and $\rho_2$, respectively, the mass conservation can only be satisfied if the liquid phase moves with a velocity $V_F$ given by $\rho_1 ds/dt = \rho_2[(ds/dt) - V_F]$, $x = s(t)$, where $s(t)$ denotes the moving phase-change line. In the liquid region a convective term contributes the heat-flow equation [25] and with FFT yields:

$$\frac{\partial T}{\partial t} \quad V_a \frac{\partial T}{\partial y} \quad V_F \frac{\partial T}{\partial y} \quad a_F \frac{\partial^2 T}{\partial y^2} \quad \frac{\partial \dot{q}}{\partial x} \qquad (9)$$

From eqs. (8) and (9) it is clear that the "convective" terms $V_F$ $T/$ $x$ and $-V_a$ $T$ $x$ have different physical meanings and origins. The former represents the convection contribution to the heat transfer while the latter is a mathematical artifact facilitating the mathematical solution only. In *absence of convection* ($V_F = 0$ we get exactly the model used in [17, 31, 32] to *simulate fuel convection* (sic!). In addition, this is the model proposed by Broekmann and Schecker [1].

## HBI solution and analysis

### Solution by the HBI method

The experiments known so far indicate two type of temperature profiles sketched in fig. 2b and c. The present study addresses that without CML and looks for a solution by the HBI method for three principle reasons: (1) explicit relationships between the basic model parameters; (2) scaling, and (3) asymptotic analysis of the results.

Assuming the fuel layer with a depth $y_0$ is $\delta(t)$, where $\delta(t)$ is the thermal penetration depth (see fig. 3b) and with a boundary condition $\lambda_F \partial T/\partial y = -F(t)$ the solution follows Goodman [22] the surface flux $F(t) = q'' =$ const. [20] and $F = F(y)$ will be used, that is the volumetric heat source is $\partial F/\partial x = \dot{q}_r$. Integration of eq. (8) over $\delta$ (see fig. 3b) gives eqs. (10) and (12):

$$\int_0^\delta \frac{\partial T}{\partial y}\, dy + V_a \int_0^\delta \frac{\partial T}{\partial y}\, dy = a_F \int_0^\delta \frac{\partial^2 T}{\partial y^2}\, dy + \frac{1}{\rho_F C_{pF}} \int_0^\delta \frac{\partial F}{\partial y}\, dy \qquad (10)$$

with boundary conditions:

$$\frac{\partial T}{\partial y}\bigg|_{x=0} = -\frac{1}{\lambda_F}(F - \rho_F H_v V_a), \quad \frac{\partial T}{\partial y}\bigg|_{y=\delta} = 0, \quad T|_{x=\delta} = T_\infty \qquad (11a, b, c)$$

$$\frac{d}{dt}(\theta - T_\infty \delta) + V_a(T_s - T_\infty) = \frac{a_F}{\lambda_F}(F - \rho_F H_v V_a) - \frac{F}{\rho_F C_{pF}} \qquad (12)$$

where $\theta = \int_0^\delta T dy$ is the *heat balance integral* (HBI) [22]. Further:

$$V_a \int_0^\delta \frac{\partial T}{\partial y}\, dy = V_a(T_{y=\delta} - T_{y=0}) = V_a(T_\infty - T_s) \qquad (13)$$

and in accordance with boundary conditions (11a, b, c) and $F_{y=0} = F$, and assuming $F_\delta = = 0$ we get:

$$a_F \int_0^\delta \frac{\partial^2 T}{\partial y^2}\, dy = a_F\left[\frac{\partial T}{\partial y}\bigg|_{y=\delta} - \frac{\partial T}{\partial y}\bigg|_{y=0}\right] =$$

$$= a_F\left[0 - \frac{\partial T}{\partial y}\bigg|_{y=0}\right] = \frac{a_F}{\lambda_F}(F - \rho_F H_v V_a) \qquad (14)$$

$$\frac{1}{\rho_F C_{pF}} \int_0^\delta \frac{\partial F}{\partial y}\, dy = \frac{1}{\rho_F C_{pF}}(F_{y=\delta} - F_{y=0}) = \frac{1}{\rho_F C_{pF}}(-F) \qquad (15)$$

With a quadratic temperature profile $T(y, t) = \beta_0 + \beta_1 y + \beta_2 y^2$, and boundary conditions (11a, b, c) we get $\beta_i(t)$:

$$\beta_0 = T_\infty - \frac{1}{2\lambda}\Phi, \quad \beta_1 = \frac{1}{\lambda}\Phi, \quad \beta_2 = -\frac{1}{2\lambda\beta}\Phi, \quad \Phi = F - \rho_F H_v V_a \qquad (16)$$

Substitution of eq. (16) in eq. (3a) and then in eq. (12) yields an equation for $\delta(t)$, namely:

$$\frac{\mathrm{d}\delta^2}{\mathrm{d}t} + 3V_a\delta^2 = 6a_F\left(1 - \frac{F}{\Phi}\right) \quad \text{with} \quad \delta(t) = 0, \ t = 0 \tag{17}$$

In case of fixed boundary ($V_a = 0$) and no flux at the boundary ($F = 0$) and $\Phi =$ $=$ const. eq. (17) is the Goodman's solution [22]. The solution to eq. (17) is [33]:

$$\delta^2(t) = \frac{2a_F B_F}{V_a}(1 - e^{-3V_a t}), \quad B_F = 1 - \frac{F}{\Phi} \tag{18a, b}$$

Therefore, the thermal penetration depth is $\delta(t) = (2a_F B_F/V_a)^{1/2}(1 - e^{-3V_a t})^{1/2}$ that correctly gives at $t = 0$ $\delta = 0$. The solution confirms the Goodman's estimation $\delta \sim (6a_F t)^{1/2}$ since expressing the exponents $1 - e^{-3V_a t}$ $1 - (1 - 3V_a/1!)$ we can read $\delta(t) \approx (2a_F B_F/V_a)^{1/2}(3V_a t)^{1/2} = (6a_F t)^{1/2}B_F^{1/2} = (6a_F t)^{1/2}$. Therefore, the volumetric heating affects the penetration depth by a factor ($B_F^{1/2} < 1$) and the temperature profile becomes:

$$\frac{T(y,t) - T_\infty}{T_s - T_\infty} = \frac{\Phi}{2\lambda_F(T_s - T_\infty)}\left(1 - \frac{y}{Z}\right)^2, \quad Z = \sqrt{6a_F t}\sqrt{B_F} \tag{19a, b}$$

*Scaling of the thermal penetration depth $\delta(t)$*

Let us try to check the initial assumption of the HBI that $y_0 \gg \delta(t)$. Really the time to boilover $t_{B0}$ is the only time limit we have from experiments. The water heating clearly indicates that the entire layer is thermally penetrated, so the answer is clear, but from an academic standpoint we should establish when with fuel depth decreasing in time the condition $y = \delta(t)$ will be satisfied. It might be reasonable to establish the upper limit of $y = \delta(t)|_{y = h}$ at $t = t_h$ that with scaling developed earlier yields $h = \delta(t_h) = (6a_F t_h)^{1/2}B_F^{1/2}$. This gives $t_h = (1/6)(h^2/a_F)1/B_F = t_0/6B_F$ and the time required the penetration depth to reach the fuel bottom $h$ is approximately 6 time shorter than the thermal diffusion time $t_0^h = h^2/a_h$ in absence of volumetric heat sources and moving surface. Taking into account that Fo$^e \sim O(1)$ [20], more exactly Fo$^e$ is close to 1 at $t = t_{B0}$, we might suggest that $\delta(t_h) = \delta(t_0^h/6) = h$. However, we have to test the contribution of $B_F$ to eq. (18a). From eq. (18b) we read:

$$B_F = 1 - \frac{F}{\Phi} = 1 - \frac{F}{F + \rho_F H_v V_a} = 1 - \left(1 + \frac{\rho_F H_v V_a}{F}\right)^{-1} = 1 - \left(1 + \frac{1}{H_p}\right)^{-1} \tag{20}$$

where $H_p = F/\rho_F H_v V_a = \dot{q}''/H_v \dot{m}''$ a dimensionless number defined in [20, 21] through the SBC.

This number $H_p$ [21] is a ratio of three dimensionless numbers characterizing the fuel layer heating (see tab. 1 and 2), that is $H_p = B_{SA}$Ste$/N_{DHS}$, where $N_{DHS} = V_a y_0/a_F$, Ste $=$ $= C_p(T_s - T_\infty)/H_v$ is the Stefan number, and $B_{SA} = \dot{q}'' y_0/\lambda_F(T_s - T_\infty)$ is dimensionless number analogue to convective number in case of radiation-conduction heating [21]. In fact,

**Table 1. Data summarized from various experiments and results derived in [21]\***

| Ref. | D [m] | $y_0$ [3] [mm] | $t_{B0}$ [1] [s] | $U_T$ [1] [$10^3$ mm/s] | Fo$^e$ [1] | $V_{a(average)}$ [2] [$10^2$ mm/s] | $N_{DHS}$ | $N_{VA}$ [$10^3$] |
|---|---|---|---|---|---|---|---|---|
| [6, 7]<br><br>Heating oil<br>($a_F = 0.877 \cdot 10^{-7}$ m²/s)$^c$<br><br>Ste = 1.373 | 0.15 | 19 | 945 | 20.1 | 0.22 | 0.01 [2] | 1.9 | 7.02 |
| | | 17 | 830 | 20.48 | 0.24 | | 1.7 | 5.75 |
| | | 13 | 625 | 20.8 | 0.31 | | 1.5 | 3.31 |
| | | 9 | 450 | 20.0 | 0.47 | | 1.3 | 1.58 |
| | | 7 | 340 | 20.58 | 0.59 | | 0.7 | 0.99 |
| | | 4 | 165 | 24.42 | 0.88 | | 0.4 | 0.33 |
| | | 2 | 90 | 22.22 | 1.94 | | 0.2 | 0.007 |
| | 0.23 | 17 | 710 | 23.94 | 0.21 | 0.011 [2] | 1.7 | 7.01 |
| | | 15 | 620 | 24.19 | 0.23 | | 1.5 | 4.71 |
| | | 13 | 530 | 24.52 | 0.27 | | 1.3 | 4.1 |
| | | 9 | 340 | 26.47 | 0.36 | | 0.9 | 2.02 |
| | | 4 | 125 | 32.0 | 0.66 | | 0.4 | 0.39 |
| | | 3 | 75 | 40 | 0.71 | | 0.3 | 0.21 |
| | | 2 | 30 | 66.66 | 0.64 | | 0.2 | 0.009 |
| | 0.5 | 15 | 345 | 43.47 | 0.13 | 0.017 [2] | 1.5 | 6.95 |
| | | 13 | 265 | 49.05 | 0.13 | | 1.3 | 6.04 |
| | | 11 | 190 | 57.89 | 0.13 | | 1.1 | 4.33 |
| | | 7 | 90 | 77.77 | 0.15 | | 0.7 | 2.89 |
| | | 5 | 70 | 71.42 | 0.24 | | 0.5 | 0.85 |
| | | 3 | 15 | 200 | 0.24 | | 0.3 | 0.329 |
| [10]<br><br>Arabian light crude-oil<br>($a_F = 0.679 \cdot 10^{-7}$ m²/s)$^c$<br><br>Ste = 1.703 | 0.3 | 3.5 | 612 | 57.18 | 0.003 | no data | | |
| | 0.6 | 20 | 492 | 40.65 | 0.083 | 2.33 | 6.87 | 6.87 |
| | | 69 | 942 | 73.25 | 0.013 | 3.33 | 33.87 | 33.87 |
| | 1 | 20 | 681 | 42.73 | 0.079 | 2.91 | 8.59 | 8.59 |
| | | 40 | 978 | 40.89 | 0.041 | 3.66 | 21.6 | 21.6 |
| | | 60 | 1310 | 45.87 | 0.024 | 4.0 | 35.34 | 35.34 |
| | | 100 | 1926 | 51.92 | 0.013 | 3.66 | 54 | 54.0 |
| | 2 | 20 | 411 | 48.66 | 0.069 | 3.08 | 9.08 | 9.08 |
| | 3.5 [4] | 27 | 402 | 67.16 | 0.037 | 3.83 | 15.24 | 15.24 |

\* These data shows experimental conditions and order of magnitude of dimensionless groups appearing in the solutions. No convection mixing layer (CML) formation in the burning fuel

Notes: (1) – calculated in [21]; (2) – calculated in [21] with data recovered from various papers of Garo *et al.*; (3) – from [6]; (4) – the diameter of the circular pan with the same area, while the real square pan is 2.7 × 2.7 m pan; $U_T$ and $R_{AV}$ data of Koseki experiments are from the original work [10], but recalculated in mm/s

**Table 2. Treatment of the data of Arai *et al.* [9] concerning the boilover onset**
**($D = 0.048$ m; $y_0 = 10$ mm)**

| Fuel | $a_F^{(1)}$ [$10^7$ m²/s] | $T_i$ [K] | $T_b^{(1)}$ [K] | $V_{a(average)}^{(1)}$ [$10^5$ m/s] | Fo$^{c\,(2)}$ | Ste | $N_{DHS}$ |
|---|---|---|---|---|---|---|---|
| Toluene | 1.03 | 293 | 383 | 1.35 | 0.43 | 0.462 | 1.35 |
| | | 318 | | | 0.29 | 0.332 | |
| | | 323 | | | 0.42 | 0.305 | |
| | | 325 | | | 0.45 | 0.295 | |
| | | 341 | | | 0.74 | 0.214 | |
| | | 355 | | | 1.01 | 0.143 | |
| Ethyl benzene | 0.88 | 293 | 409 | 1.5 | 0.45 | 0.558 | 1.69 |
| | | 323 | | | 0.24 | 0.432 | |
| | | 330 | | | 0.18 | 0.398 | |
| | | 345 | | | 0.08 | 0.322 | |
| n-Decane | 0.753 | 293 | 433 | 1.19 | 0.34 | 1.149 | 1.58 |
| | | 323 | | | 0.22 | 0.892 | |
| | | 348 | | | 0.11 | 0.694 | |
| | | 355 | | | 0.10 | 0.636 | |
| | | 362 | | | 0.07 | 0.581 | |

Notes: (1) – The paper of Garo *et al.* [6] was used as a source for some average values of the fuel properties due to deficiences in the original paper [21]; (2) – from [21]. $N_{VA}$ was calculated with approximately constant flame temperature of about 1100 K typical for hydrocarbon pool fire through eq. (2)

$B_{SA}$ is a ratio of two length scales, $B_{SA} = y_0/Z_0$, where $Z_0 = \dot{q}''/\lambda_F(T_s - T_\infty)$. The estimates in [21] reveal that $H_p \ll 1$ that allows to $1/H_p \gg 1$ and therefore $B_F \gg 1$. Hence, from eq. (20) we have $t_h \approx t_0/6$ and there is no effect of the radiation flux. The latter implies that with internal heat generation the temperature profile [22] is $T(\delta,t) = (Q/\rho_F C_{pF})[1 - (1 - y/\delta)^3]$ that resembles the present solution (19). In this case the penetration depth is $\delta = [(24a_F \int_0^t Q^2 \cdot dt)Q^{-2}]^{1/2}$ where $Q = \int_0^t \dot{q}''(t)dt$ with non-time varying surface flux, $\dot{q}''(t) = F(t) = $ const. In view of $Q = Ft$ we can read $\delta = [24a_F(Q^2t/Q^2)]^{1/2} = (24a_Ft)^{1/2}$. Further, we read $t_h = t_0^h/24$ that is shorter than $t_h \approx t_0^h/6$. Therefore, with a certain process approximation we may assume that at the beginning $\delta = y_0 = h$ that simplify the solution, *i. e.* the HBI should be applied over the entire fuel layer thickness $h$. Such an example is developed next.

## Alternative HBI solution in ablation approximation with $\delta > h$

*Linear boundary conditions*

Equation (5) under stationary conditions with $y = 0$, $T = T_s$, and $x = \infty$, $T = T_\infty$, and $dT/dy = 0$ at $y = \delta = h$ has a solution expressed as (7). The transient heat transfer solution can be expressed as [34]:

$$T - T_\infty = (T_s - T_\infty) e^{-\frac{V_a y}{a_F}} D(t) \tag{21}$$

The time-dependent function $D(t)$ can be determined by HBI with $\delta(t) > y_0 = h$, namely:

$$\int_0^h \frac{dD(t)}{dt}(T_s - T_\infty) e^{-\frac{V_a y}{a_F}} dy = a_F(T_s - T_\infty) \int_0^h \left(\frac{V_a}{a_F}\right)^2 e^{-\frac{V_a y}{a_F}} D(t) dy \tag{22}$$

$$\frac{dD(t)}{dt} = -\frac{V_a^2}{a_F} D(t) \implies D(t) = D_0 e^{-\frac{V_a^2}{a_F} t} \tag{23}$$

The initial condition $D_0 = D(0)$ could be determined from the temperature at the flaming surface ($y = 0$ at $t = 0$) that rises immediately to $T = T_s$. This condition provides $D_0 = D(0) = 1$ and we read:

$$\frac{T - T_\infty}{T_s - T_\infty} = e^{-\frac{V_a y}{a_F}} e^{-\frac{V_a^2}{a_F} t} \tag{24}$$

Note: The approach to determine $D_0 = D(0)$ differs from that suggested by Zubarev [34] where the condition $T_\infty = (1/h) \int_0^h T(y,t) dy$ at $y = 0$ at $t = 0$ is used.

*Non-linear (Stefan) boundary condition*

The general time-dependent solution to eq. (5) with $y \to \infty$, $T = T_\infty$ and $y = 0$, $- \partial T / \partial x = = (1/\lambda_F)(F + \rho_F C_{pF} H_v) = \Phi/\lambda_F$ similar to eq. (21) is:

$$T - T_\infty = \Phi \frac{a_F}{V_a} e^{-\frac{V_a}{a_F} y} P(t) \tag{25}$$

Application of HBI about $P(t)$ just like eq. (23) yields:

$$\frac{dP(t)}{dt} + \frac{V_a^2}{a_F} P(t) = 0 \implies P(t) = P_0 e^{-\frac{V_a^2}{a_F} t} \tag{26}$$

The Stefan boundary condition $- \partial T / \partial y = \Phi/\lambda_F$ at $y = 0$ simply provides $P_0 = = P(0) = 1$, that leads to:

$$\frac{T - T_\infty}{T_s - T_\infty} = \frac{\Phi}{\lambda_F (T_s - T)_\infty} \frac{a_F}{V_a} e^{-\frac{V_a}{a_F} y} e^{-\frac{V_a^2}{a_F} t} \tag{27}$$

*Analysis of the time constant $V_a^2 / a_F$*

The exponent $(V_a^2/a_F)t$ in eqs. (26) and (27) can be rearranged as:

$$\frac{V_a^2}{a_F}t \quad \frac{V_a^2}{h^2}\frac{h^2}{a_F}t \quad \left(\frac{V_a}{h}\right)^2 t_0 t \quad \left(\frac{V_a}{h}\right)^2 t_0^2 Fo^h \quad \left(\frac{t_0}{\tau_0}\right)^2 Fo^h \tag{28}$$

Here $t_0^h = h^2/a_F$ is the thermal diffusivity time scale and $Fo^h = t/t_0^h = h^2 t/a_F$. With $h_{max} = y_0$ (at $t = 0$) as a length scale [21] the Fourier number is $Fo = y_0^2 t/a_F$ and $t_0 = y_0^2/a_F$. The theoretical time required for complete combustion of the fuel $\tau_0 = h/V_a$.

Data about $t_{B0}$ and $Fo^e = t_{B0}/t_0$ are available elsewhere [20, 21]. Thus, the exponent of $D(t)$ at $t = t_{B0}$ is $(V_a^2/a_F)t_{B0} = (t_0/\tau_0)^2 Fo^e = V_a^2/U_0 U_T$. Here $U_0 = y_0/t_0$ and $U_T = y_0/t_{B0}$ (Koseki's wave) are two hypothetical velocities defined through macroscopic variables. Therefore we can express $(V_a^2/a_F)t$ at $t = t_{B0}$ as $(V_a^2/a_F)t_{B0} = N_{DHS}(V_a/U_T)$. Consequently, the profile (27) reads at:

$$\Theta_{B0} \quad \frac{T_{B0} \quad T_\infty}{T_s \quad T_\infty} \quad \frac{\Phi}{\lambda_F(T_s \quad T_\infty)}\frac{a_F}{V_a}e^{\frac{V_a y}{a_F}}e^{(N_{DHS})^2 Fo^e} \tag{29}$$

## Boilover onset – dimensionless scaling of the semi-analytical solution of SAM

With the profile (29), data about $Fo^e$ and the location of the water $y = y_F$ we may define the temperature at the fuel-water (F/W) interface $T|_{y \quad y_w} = T_{B0}$. This case formulates a direct A problem. Alternatively, definition of $Fo^e$ through (29) with known $T = T_{B0}$ formulates an inverse B problem.

- **A problem – Time to boilover $t_{B0}$.** Simple rearrangement of expression (29) at $t = t_{B0}$ yields:

$$A_{B0} \quad \Theta_{B0} \quad \left(\frac{\Phi}{\lambda_F(T_s \quad T)_\infty}\frac{a_F}{V_a}e^{\frac{V_a y}{a_F}}\right)^{1} \quad e^{(N_{DHS})^2 Fo^e} \tag{30}$$

$$Fo^e \quad \frac{\ln A_{B0}}{N_{DHS}^2} \quad t_{B0} \quad \frac{\ln A_{B0}}{N_{DHS}^2}\frac{y_0^2}{a_F} \tag{31}$$

- **B problem – Temperature $T = T_{B0}$ at F/W interface.** From (29) we get:

$$\Theta_{B0} \quad B_{B0}e^{[(N_{DHS})^2 Fo^e]} \tag{32}$$

with

$$B_{B0} \quad \frac{\Phi}{\lambda_F(T_s \quad T)_\infty}\frac{a_F}{V_a}e^{\frac{V_a y}{a_F}} \quad N_0 \quad 1 \quad \frac{1}{H_p} \quad \delta_F \tag{33}$$

Here $N_0 = Fy_0/\lambda_F(T_s - T_\infty)$ is the radiation-conduction parameter [20]; $\delta_F = y_F/y_0$ with $y_F/y_0 + y_w/y_0 = 1$. Equations (30-33) call for scaling analyses addressing $A_{B0}$, $B_{B0}$, and $Fo^e$.

*Dimensionless scaling of $Fo^e$, $A_{B0}$, and $B_{B0}$*

Simple rearrangement of the nominator of eqs. (32) or (33) with approximation of the exponent in a series (valid for small $N_{DHS}\delta_F$) since $N_{DHS} \ll 1$ (see [20]) and $\delta_F = y_F/h < 1$ provides:

$$A_{B0} \approx \Theta_{B0}\frac{\Phi y_0}{\lambda_F(T_s-T_\infty)_\infty}\frac{1}{N_{DHS}}N_{DHS}\delta_F \approx \Theta_{B0}\frac{\Phi y_0}{\lambda_F(T_s-T_\infty)_\infty}\delta_F \quad \text{(34a, b)}$$

Since $\Phi = F + \rho_F H_v V_a$ (see eq. (20)) we read from eq. (34) that $\Phi y_0/\lambda_F(T_s-T_\infty) = N_0 + N_{DHS}/\mathrm{Ste}$ and:

$$A_{B0} \approx \Theta_{B0}\left(N_0 + 1\frac{B_{SA}N_{DHS}}{\mathrm{Ste}}\right)\delta_F \quad A_{B0} \approx \frac{\Theta_{B0}}{N_0\delta_F}H_p \quad (35)$$

since from eq. [21]:

$$H_p \ll 1 \quad 1 + \frac{1}{H_p} \quad H_p \quad \frac{B_{SA}N_{DHS}}{\mathrm{Ste}}\,[21] \quad (36)$$

Now, taking into account that $\ln x \approx x - 1$ with $0 < x \le 2$, eq. (31a) scales as:

$$Fo^e \approx \ln\frac{\dfrac{\Theta_{B0}H_p}{N_0\delta_F}}{N_{DHS}^2} \quad Fo^e \approx \frac{\Theta_{B0}}{N_0\delta_F}H_p\left(1+\frac{1}{N_{DHS}^2}\right) \quad (37)$$

In view of the orders of magnitude of the dimensionless numbers, *i. e.*, $\Theta_{B0} \sim O(1)$, $\delta_F \sim O(1)$, $N_0 \sim O(1)$ for laboratory fire and $N_0 \sim O(10^2)$ for large fire [21] and $N_{DHS} \sim O(1)$ we get $Fo^e \sim H_p/N_{DHS}^2$ that confirms the scaling relationship $Fo^e \sim (\Theta_{B0}^2/\mathrm{Bu})H_p^m$ established in [21] through a physical analysis only. Further, in explicit (dimensional) form the last scaling relationship gives $t_{B0} \sim y_0^2 H_p/a_F(V_a y_0/a_F)^{-2} \sim (a_F/V_a^2)(F/H_v\dot m) \quad D^{1/2} \quad t_{B0} \sim D^{1/2}$ since $F \sim D^{1/2}$ (see eq. 2). The last scaling $t_{B0} \sim D^{1/2}$, established in [20] and confirmed here, implies that both heat conduction and radiation flux absorption contribute the heat budget.

From eq. (32), in view that $N_{DHS} \sim O(1)$ [21] (see tab. 1 and 2 too) and with $Fo^e \ll 1$ [21], we get $e^{N_{DHS}^2 Fo^e} \approx 1 + N_{DHS}^2 Fo^e$. This allows the fuel/water critical temperature to be expressed as $\Theta_{B0} \approx N_0 H_p \delta_F (1 - N_{DHS}^2 Fo^e)$. This is a reasonable since from [21] we know that $N_0 \sim O(1)$. Hence we have $\Theta_{B0} \sim O(1)$ or more precisely $\Theta_{B0} < 1$. The estimate $\Theta_{B0} \sim O(1)$ need a definition of the water temperature at $t = t_{B0}$. For instance, for pure water the superheat temperature is about $T_{W-SH} = 270\ °C$ in view of *homogenous nucleation* as explosion onset [33, 34]. In fuel storage tanks the bottom water is mainly due to condensate so impurities of any kind exist and can reduce the water superheat temperature less than $T_{W-SH} = 270\ °C$. Garo *et al.* [6, 7] report $T_{B0} = 373\ K$ and $T_\infty = 20\ °C$ that gives $\Theta_{B0} = 0.335$ [21]. The experiments of Koseki *et al.* [10] provide $\Theta_{B0} = 0.432$ [21]. These numerical values will be used in the wave analysis performed next.

## Wave-like analysis of the semi-analytical solutions

*Thermal penetration depth and wave-like solutions*

The geometrical terms of eqs. (24) and (27) define a *heat penetration depth* $y_p$ as a distance where the surface temperature is attenuated $e$ times that *differs physically* from the *thermal penetration depth* $\delta(t)$ used by HBI method. Hence, from eq. (24) we have $\Theta = (T - T_\infty)/(T_s - T_\infty) = e^{-V_a y/a_F}$ that through $\ln\Theta = \ln(1/e) = -V_a y/a_F$ defines the heat penetration depth $y_{p0} = a_F/V_a$. With a time-dependent term we get $y_{pt} = a_F/V_a + V_a t$. Similarly from eq. (27) we get:

$$y_{p0\text{-non-linear}} = \frac{a_F}{V_a}\left[1 + \ln\left(\frac{a_F}{V_a}\frac{\Phi}{\lambda_F(T_s - T_\infty)}\right)\right] \tag{38}$$

The term in the square brackets of eq. (38) is dimensionless, closes unity and from eq. (38) we have $y_{p0\text{-non-linear}} = (a_F/V_a)\{1 + \ln[(a_F/V_a)\Phi/\lambda_F(T_s - T_\infty)]\} = (a_F/V_a)(1 + \ln Y) \approx a_F/V_a$. Since $Y \sim O(1)$ because $Y = [\Phi y_0/\lambda_F(T_s - T_\infty)](1/y_0) = N_{VA}/y_0 \approx O(1)$ (see tab. 1 and 2). For most of fuels (tab. 1 and 2) we have $a_F/V \approx 8$ mm. For fuels with $y_0 > 8$ mm we have $N_{VA} \approx (3-8)10^{-3}$. Both eqs.(24) and (27) can be represented as attenuated heat pulses propagating downward to the tank bottom with a velocity $V_a$, namely:

$$\Theta \approx e^{-\frac{V_a}{a_F}(y - V_a t)} \quad \text{and} \quad \Theta \approx \frac{\Phi}{\lambda_F(T_s - T)_\infty}\frac{a_F}{V_a}e^{-\frac{V_a}{a_F}(y - V_a t)} \tag{39a, b}$$

The pulse magnitude (see eq. 39b) depends on the dimensionless number $N_p = [\Phi/\lambda_F(T_s - T_\infty)](a_F/V_a) = \Phi y_{p0}/\lambda_F(T_s - T)_\infty$ similar to the radiation conduction number $N_0$ [21]. Further, $N_p = N_0 \tilde{y}_{p0}$ with $\tilde{y}_{p0} = a_F/V_a y_0 \approx 1/N_{DHS}$ that is $N_p = N_0/N_{DHS}$.

*Koseki's heat wave*

Koseki [19] defines experimentally a thermal wave with velocity $U_T = y_0/t_{B0}$. The values of $\Theta_{B0}$ commented previously gives $\ln\Theta_{B0(Garo)} = \ln 0.355 = -1.093$ and $\ln\Theta_{B0(Koseki)} = \ln 0.432 = -0.839$ that are very close to $\ln(1/e) = -1$ defining the thermal penetration depth $y_p$. For seek of simplicity we assume that at $t = t_{B0}$ the left-hand side of both eqs. (39) is $\Theta_{B0} = 1/e$ that yields:

$$\ln\Theta_{B0} = 1 = \frac{y_0}{y_{p0}} - \tilde{y}_F \frac{V_a}{y_0}t_{B0} \quad \frac{U_T}{V_a} = \frac{y_0}{\tilde{y}_F y_0 - y_{p0}} \quad U_T \sim y_0 \tag{40a, b, c}$$

Here $\tilde{y}_F = y_F/y_0$ is the dimensionless fuel depth and $\tilde{y}_w = y_w/y_0 = 1 - \tilde{y}_F$ where $1/(\tilde{y}_F y_0 - y_{p0}) > 0$. Estimations (40b, c) confirms that $U_T$ is linearly proportional to the fuel depth, $U_T \sim V_a$, which is the higher burning rates, the shorter pre-boilover times. Similarly in case of SBC we get $U_T/V_a = y_0/[\tilde{y}_F y_0 - y_{p0}\ln(N_0/N_{DHS})] \quad U_T \sim y_0$. Then, eqs.32 and 33 provide:

$$\text{Fo}^e \quad \frac{t_{B0}}{t_0} \quad \frac{1}{N_{DHS}} \quad \widetilde{y}_F \quad \frac{1}{N_{DHS}} \frac{\ln N_p}{N_{DHS}} \qquad \text{Fo}^e \quad \frac{1}{N_{DHS}} \frac{y_F}{y_0} \qquad (41a, b)$$

Bearing in mind the Bouguer number $Bu = \mu y_0$ and that $H_p \quad N_{DHS}$ (see eq. 36) we get from eqs. (41b, c) a power-law $\text{Fo}^e \sim H_p^m/Bu$ [21]. For instance, the data of Koseki [10] scaled in [21] reveal that $\text{Fo}^e \sim 1/H_p \quad 1/N_{DHS}$ that gives immediately eqs. (41a, b).

## Simplified estimation of the time to boilover

### Heat conduction dominating heating

The penetration depth concept and the assumption $\ln(1/e) \quad -1$ allow suggesting that at $t = t_{B0}$ the temperature at the F/W interface becomes $\Theta \quad e$. Thus, since the fuel depth decreases in time, we might suggest that at $t = t_{B0}$, we get $y_{F(t \quad t_{B0})} = y_p \quad a_F/V_a$ that gives $y_{F(t \quad t_{B0})} = y_p = y_0 - V_a t_{B0}$ and $t_{B0} = (y_0 - y_p)/V_a = y_0/V_a - a_F/V_a^2 = \tau_0 - a_F/V_a^2$. Besides, with a simple algebra we get $t_{B0} = \tau_0(1 - 1/N_{DHS})$ that is in fact the estimation $t_{B0} \sim 1/N_{DHS}$ we got earlier. Additionally, taking into account that eq. (36), that is, $H_p = B_{SA}N_{DHS}/\text{Ste}$ we get:

$$t_{B0} \quad \tau_0 \ 1 \quad \frac{1}{H_p}\frac{\text{Ste}}{B_{SA}} \qquad t_{B0} \quad \frac{\tau_0}{2} \ 1 \quad \frac{1}{F} \qquad t_{B0}$$
$$y_0 \frac{1}{F} \qquad t_{B0} \quad y_0 \frac{1}{\sqrt{D}} \qquad t_{B0} \quad \frac{y_0}{D}\sqrt{D} \qquad (42)$$

that confirms estimations derived in [20, 21] through scaling of experimental data. The last expression of eq. (42) defines the Koseki's heat wave velocities, for heat conduction dominating regime, as $U_T = y_0/t_{B0} = V_a[N_{DHS}/(N_{DHS} - 1)]$ upon the condition $N_{DHS} > 1$.

### Critical fuel thickness

The requirement $N_{DHS} > 1$ (*i. e.* $y_0 > a_F/V_a$) applied to the heating oil used by Garo *et al.* [6] defines $y_0 \quad 8 \cdot 10^{-3}$ m (see data with $N_{DHS} \quad 1$ in tab. 1). The same test applied to the fuels of Arrai *et al.* [9] listed in tab. 2 provides as follows: $y_0 \quad 7.6$ mm for toluene; $y_0 \quad 5.8$ mm for ethyl benzene; and $y_0 \quad 6.3$ mm for n-Decane. These experiments are generally defined in the literature as *thin-layer boilover* on the basis of the initial fuel layer thickness. Now, we define a principle thermophysical *criterion distinguishing thin and thic*k layer boilover. The same numerical test applied to the data of Koseki *et al.* [10] with a mean burning velocity of about $3 \cdot 10^{-5}$ m/s generally define the bulk of his experiments as thick-layer boilover.

Calculations of $t_{B0}$ as $t_{B0} = \tau_0(1 - 1/N_{DHS})$ and through eq. (42) are summarized in tab. 3. The calculations are rough but reveal that the predicted $t_{B0}$ approaches the experimentally defined values when $N_{DHS} \quad 1.3$ for the fuel of Garo [6]. In the experiment of Koseki *et al.* [10] assuming $0.7 \quad t_{B0(experimental)}/t_{B0}$ (eq. 42) $\quad 1.2$ the range of variations of

**Table 3. Time to boilover experimental and predicted values with auxiliary information enable calculations thorugh eqs. (42), (47c), and (48a, b)**

| Ref. | D [m] | $y_0$ [mm] | $t_0$ [s] | $\dot{v}_{(average)}$ [$10^3$ mm/s] | $N_{bois}$ | Bu | $t_{60}$ (1) [s] exp. | $t_{60}$ [s] eq. 42 | $t_{bois,+}/t_{bois}$ (eq. 42) | $t_{60}$ [s] eq. (47c) | $t_{bois,+}/t_{bois}$ (eq. 47c) | $t_{60}$ [s] eqs. (48a, b) | $t_{bois,+}/t_{bois}$ (eq. 48a, b) |
|---|---|---|---|---|---|---|---|---|---|---|---|---|---|
| [6, 7] Heating oil ($\alpha_f = 0.877 \cdot 10^{-7}$ m²/s) (3), Ste = 1.373, $\frac{d_s}{V_s} = 8$ mm, $y_0 =$ | 0.15 | 19 | 1900 | 0.01 (2) | 1.9 | 4.97 | 945 | 900 | 1.05 | | | | |
| | | 17 | 1700 | | 1.7 | 4.45 | 830 | 700 | 1.18 | | | | |
| | | 13 | 1300 | | 1.5 | 3.40 | 625 | 433 | 1.44 | | | | |
| | | 9 | 900 | | 1.3 | 2.36 | 450 | 207 | 2.17 | | | | |
| | | 7 | 700 | | 0.7 | 1.83 | 340 | | | 83 | 4.09 | 99 | 3.43 |
| | | 4 | 400 | | 0.4 | 1.05 | 165 | | | 82 | 2.02 | 97.6 | 1.7 |
| | | 2 | 200 | | 0.2 | 0.52 | 90 | | | 84 | 1.07 | 100 | 0.9 |
| $\mu = 262$ m⁻¹ [6, 7, 21] | 0.23 | 17 | 1545 | 0.011 (2) | 1.7 | 4.45 | 710 | 636 | 1.11 | | | | |
| | | 15 | 1363 | | 1.5 | 3.93 | 620 | 500 | 1.24 | | | | |
| | | 13 | 1181 | | 1.3 | 3.40 | 530 | 272.7 | 1.94 | | | | |
| | | 9 | 818 | | 0.9 | 2.36 | 340 | | | 75 | 4.53 | 89.2 | 3.81 |
| | | 4 | 363 | | 0.4 | 1.05 | 125 | | | 75 | 1.66 | 89.2 | 1.4 |
| | | 3 | 273 | | 0.3 | 0.786 | 75 | | | 75.56 | 0.99 | 84.5 | 0.88 |
| | | 2 | 182 | | 0.2 | 0.52 | 30 | | | 76.73 | 0.39 | 91.1 | 0.33 |
| | 0.5 | 15 | 882 | 0.017 (2) | 1.5 | 3.93 | 345 | 294 | 1.17 | | | | |
| | | 13 | 765 | | 1.3 | 3.40 | 265 | 176.4 | 1.5 | | | | |
| | | 11 | 647 | | 1.1 | 2.88 | 190 | 58.8 | 3.23 | | | | |
| | | 7 | 412 | | 0.7 | 1.83 | 90 | | | 72.03 | 1.25 | 83 | 1.07 |
| | | 5 | 294 | | 0.5 | 1.31 | 70 | | | 48.8 | 0.97 | 58.07 | 1.2 |
| | | 3 | 176 | | 0.3 | 0.786 | 15 | | | 48.8 | 0.30 | 58.07 | 0.31 |
| [10] Arabian light crude-oil ($\alpha_f = 0.679 \cdot 10^{-7}$ m²/s) (3), Ste = 1.703, $\frac{d_s}{V_s} = 2.26$ m, $y_0 =$ | 0.3 | 20 | 858 | no data | 5.89 | | 612 | 145 | 0.617 | | | | |
| | 0.6 | 69 | 2072 | 2.33 | 6.87 | | 492 | 733 | 0.468 | | | | |
| | | 20 | 687 | 3.33 | 33.87 | | 942 | 2010 | 1.11 | | | | |
| | 1 | 40 | 1092 | 2.91 | 8.59 | | 681 | 607.27 | 0.938 | | | | |
| | | 60 | 1500 | 3.66 | 21.6 | | 978 | 1042 | 0.899 | | | | |
| | | 100 | 2732 | 4.0 | 35.34 | | 1310 | 1457 | 0.718 | | | | |
| | 2 | 20 | 649 | 3.66 | 9.08 | | 1926 | 2681 | 0.711 | | | | |
| | 3.5 (4) | 27 | 705 | 3.08 | 15.24 | | 411 | 577.836 | 0.610 | | | | |
| | | | | 3.83 | | | 402 | 658.703 | | | | | |

Notes: (1) – calculated in [21]; (2) – calculated in [21] with data recovered from various papers of Garo et al.; (3) – from [6]; (4) – the diameter of the circular pan with the same area, while the real square pan is 2.7 2.7 m pan; $U_f$ and $R_{AV}$ data of Koseki experiments are from the original work [10], but recalculated in mm/s

$N_{DHS}$ is 8 ≤ $N_{DHS}$ ≤ 54. More strong conditions of 0.9 ≤ $t_{B0(experimental)}/t_{B0}$ (eq. 42) ≤ 1.1 defines a range of 8 ≤ $N_{DHS}$ ≤ 21. Theses discrepancies call for attention on the change of the fuel heating mechanism. The proposed empirical method provides reasonable times to boilover for thick fuel layers defined by the condition $y_0 \gg a_F/V_a$. In fact $y_0 \gg a_F/V_a$ defines the minimum fuel thickness beyond which the 1-D heat conduction equation is applicable. The latter implies that with $y_0 \ll a_F/V_a$ the entire bed is heated volumetrically by the absorbed flame radiation with negligible temperature gradients.

*Simplified estimation of the time to*
*boilover – dominating radiation heating*

Neglecting all terms with temperature gradients in eq. (9) yields a kinetic equation:

$$\frac{\partial T}{\partial t} = \frac{\dot{q}_s \mu}{\rho_F C_{pF}} e^{-\mu y} \rightarrow \Theta = \frac{N_0}{Bu} Fo_3^2 = \frac{N_{DHS}}{N_0} Fo_3 \qquad (43a, b)$$

with a length scale $L_0 = \lambda_F(T_s - T_\infty)/\dot{q}_s$, time scale $\tau = L_0^2/a_F$, $\Theta = (T - T_\infty)/(T_s - T_\infty)$, and $Fo_3 = t/\tau = t a_F/L_0^2$ and $T = T_\infty$ at $t = 0$.

The condition to neglect the pseudo-convective term $(V_a \rho_F C_{pF}) \partial T/\partial y)$ in eq. (9) [21] with a length scale $L_0$ is $V_a L_0/a_F \ll 1 \rightarrow a_F/V_a \gg L_0 \rightarrow V_a y_0/a_F \ll y_0/L_0 \rightarrow N_{DHS} \ll N_0$. Further with dominating source term $\sim \partial q_r/\partial y$ its dimensionless coefficient has to be of order of magnitude 1, that is $L_0 \approx 1/\mu \rightarrow \lambda_F(T_s - T_\infty)/\dot{q}_s y_0 \approx 1/\mu y_0 \rightarrow N_0 \approx Bu$. Since $Bu$ can be calculated before the experiments (see tabs. 1-3) we get $N_{DHS} < Bu$ as a final estimate. From tab. 3 with average value of Bu about 1-1.3 we get $N_{DHS} < 1 \rightarrow y_0 \ll a_F/V_a$ as a final estimate. From tab. 3 with average value of Bu about 1-1.3 and we get $N_{DHS}$ that gives $y_0 \approx 8$ mm. Hence, the radiation absorption as a heat transfer mechanism dominates within layers with $y_0 \approx 8$ mm. Further, with $\mu y \ll 1$ a truncated series approximation of the exponent in eq. (43a) yields:

$$\frac{\partial T}{\partial t} = \frac{\dot{q}_s \mu}{\rho_F C_{pF}} e^{-\mu y} \rightarrow \frac{\partial T}{\partial t} = \frac{\dot{q}_s \mu}{\rho_F C_{pF}}(1 - \mu y) \rightarrow \frac{\partial \Theta}{\partial t^*} = \frac{1}{\mu L_0} - y^* \qquad (44)$$

The dimensionless temperature profile is $\Theta = (1/\mu L_0 - y^*)t^*$ with $y^* = y/L_0$ and $\Theta = 0$ at $t = 0$. At the boilover onset $t^*_{boilover} = t_{B0}a_F/L_0^2$ and accordance with the comments above $\Theta \approx 0.335$ in the experiments of Garo *et al.* [6, 9]. The temperature profile of eq. (44) gives:

$$t^* = \frac{\Theta_{B0}}{1/\mu L_0 - y^*} \rightarrow t_{B0} = \frac{\Theta_{B0}}{1/\mu L_0 - y^*} \frac{L_0^2}{a_F} \qquad (45a, b)$$

Further, the condition $L_0 \approx 1/\mu \rightarrow N_0 \approx Bu$ simplifies the solution and with $y_0/L_0 = = N_0 \approx Bu$ we read:

$$t_{B0} = \Theta_{B0} \frac{1}{1 - \dfrac{y}{y_0}} \left(\frac{L_0^2}{y_0^2}\right)^2 \frac{y_0^2}{a_F} \rightarrow t_{B0} = \frac{0.335}{0.4}\left(\frac{1}{\mu y_0}\right)^2 \frac{y_0^2}{a_F} \qquad (46a, b)$$

The relative thickness of the residual fuel layer is about $\Delta/y_0 = (1 - y)/y_0 \approx 0.4$ as an average value estimated from the experiments of Garo *et al.* [6, 9] and calculated in [21] that is $1 - y/y_0 \approx 0.6$:

$$t_{B0} \approx 0.84 \left(\frac{1}{Bu}\right)^2 \frac{y_0^2}{a_F} \quad t_{B0} \approx 0.84 \left(\frac{1}{Bu}\right)^2 \frac{V_a y_0}{a_F} \frac{y_0}{V_a} \quad 0.84 \left(\frac{1}{Bu}\right)^2 N_{DHS} \tau_0 \quad (47a, b, c)$$

The numerical pre-factor in eq. (47) $0.84 \approx 1$ and we may assume it, *by convention*, as equal to unity. Hence, we get approximately:

$$t_{B0} \approx \left(\frac{1}{Bu}\right)^2 \frac{y_0^2}{a_F} \quad \text{or} \quad t_{B0} \approx \left(\frac{1}{Bu}\right)^2 N_{DHS} \tau_0 \qquad (48a, b)$$

In accordance with (47c) the time to boilover is completely predictable by macroscopic variable known before the experiments with the only discriminating condition $N_{DHS} < 1$ (*i. e.* $y_0 < a_F/V_a$). The predicted values of a $t_{B0}$ by eq. (47c) and eqs. (48a, b) are summarized in tab. 3. In fact eqs. (47b, c) defines the Koseki's heat wave velocity in dominating radiation regime, namely: $U_T = y_0/t_{B0} = 1.2 \, Bu^2 (a_F/y_0)$ and $U_T = y_0/t_{B0} = 1.2 Bu^2 (V_a/N_{DHS})$.

The calculated values strongly reveal that around the critical fuel thickness defined by $y_0 = a_F/V_a$ there is a zone of large discrepancy where both heat conduction and radiation absorption take place in equal order of magnitude. As the fuel layer grows toward thicker to depths approximately $(2-2.5)(a_F/V_a)$ the ratio $T_{B0(\exp)}/t_{B0(eq. 42)}$ approaches 1. Similarly in case of radiation dominating regime the ratios $T_{B0(\exp)}/t_{B0(eq. 47c)}$ or $T_{B0(\exp)}/t_{B0(eq. 48a,b)}$ are about 1 with fuel layers of depths about $0.5(a_F/V_a)$. These approximate limits figure the width of the zone where both heat transfer mechanism take place and formulates a new problem beyond the scope of the present work.

## Conclusions

HBI method and scaling analysis of both the initial governing equations and the solutions were performed thoroughly through the text development but some principle issues can be outlined again.

(1) The initial analysis formulated two principle Stefan problems pertinent to pool fire boilover. The inverse Stefan problem only in absence of CML simplifies the analysis and allows demonstrating the power of the HBI method. The closed form solutions allow scaling and provide results directly applicable to the time to boilover.

(2) The wave-like forms of the solutions permits to evaluate the Koseki's heat wave velocity as a part of the rate controlling group pertinent to evolution of the temperature profiles across the burning fuel layer. Koseki's heat wave velocity can be expressed through dimensionless independent variables known before the experiment that is in fact a theoretical derivation of this parameter pertinent to boilover and introduced intuitively in [10].

(3) The approximate analysis based on the heat penetration depth led to reasonable results in calculations of the time to boilover in two distinguished heating regimes: dominating heat conduction and dominating heat radiation absorption. The principle results are the critical fuel depth $y_0 = a_F/V_a$ distinguishing the thin-layer boilover and thick-layer counterparts. It is based on fuel physical properties pertinent to their heat (heat diffusivity) and mass transfer characteristics (burning rate) and avoid any geometrically based definitions.

## Acknowledgments

The work was supported by the author's home university (UCTM) through a research grant NN2/60-2007.

## Nomenclature

$a_F$   – thermal diffusivity of fuel, [m²s⁻¹]

$a_F$   – thermal diffusivity of fuel, $[\text{m}^2\text{s}^{-1}]$
$B_F$   – dimensionless flux defined by eq. (18)
$C_p$   – specific heat of flame gases, $[\text{Jkg}^{-1}\text{K}]$
$C_{pF}$   – heat capacity of fuel, $[\text{Jkg}^{-1}\text{K}]$
$D$   – pool (tank) diameter, [m]
$F$   – surface density of the flux, $[\text{Wm}^{-2}\text{s}^{-1}]$ (a symbol used to replace $\dot{q}$ in the formulae for clarity of expression)
g   – gravity acceleration, $[\text{ms}^{-2}]$
$H_v$   – the latent heat of vaporization, [Jkg]
$h$   – fuel depth (in coordinates assumed $h = y_s$), [m]
$\dot{m}$   – mass burning rate per unit surface area of the pool ($\dot{m} = \rho_F V_a$), $[\text{kgm}^{-2}\text{s}^{-1}]$
$Q$   – total heat accumulated by the layer over a limited time interval, [W]
$\dot{q}_c$   – heat conduction flux from the interface toward the fuel depth, $[\text{Wm}^2\text{s}^{-1}]$
$\dot{q}_r$   – volumetric heat sources (eqs. 4a,b), $[\text{Wm}^3\text{s}^{-1}]$
$\dot{q}_s$   – surface density of the flux, $[\text{Wm}^2\text{s}^{-1}]$
$r(t)$   – surface regression rate (in the present context, $r(t) = V_a$), $[\text{ms}^{-1}]$
$T$   – temperature, [K]
$T_f$   – flame temperature, [K]
$T_s$   – vaporization temperature of the fuel, [K]
$T_\infty$   – temperature of the environment undisturbed by the heat release from the flame, [K]
$T_\infty^f$   – temperature of the fire environments, i. e. the air, [K]
$t$   – time, [s]
$t_h$   – time required the thermal penetration layer $\delta$ to reach the bottom ($t = t_h$ at $\delta = h$), [s]
$t_0$   – time scale ($t_0 = y_0^2/a_F$), [s]
$t_{B0}$   – time of boilover onset, [s]
$y$   – vertical coordinate, [m]
$y_F$   – fuel depth, [m]
$y_p$   – heat penetration depth, [m]
$y_s$   – current position of the flaming interface moving with a velocity $V_a$, [m]
$y_w$   – water depth, [m]

$y_0$    – initial fuel-water layer depth ($y_0 = h|_{t\ 0}$), [m]
$U_T$    – Koseki's heat wave velocity expressed as $U_T = h/t_{B0}$ or $U_T = y_0/t_{B0}$, [ms$^{-1}$]
$U_0$    – hypothetic velocity ($U_0 = y_0/t_0 = a_F/y_0$), [ms$^{-1}$]
$V_a$    – velocity of the interphase interface motion (fuel regression rate), [ms$^{-1}$]
$V_F$    – convective velocity in the fuel, [ms$^{-1}$]
$V_s$    – velocity of propagation of the convection mixing layer (hot zone), [ms$^{-1}$]

*Greek letters*

$\delta$    – thermal penetration depth in accordance with the HBI assumption, [m]
$\Theta$    – dimensionless temperature
$\Theta_{B0}$    – dimensionless temperature at the boilover onset
$\theta$    – heat balance integral (HBI)
$\lambda_F$    – thermal conductivity of the fuel, [Wm$^{-1}$K$^{-1}$]
$\mu$    – effective average radiation absorption (or extinction) coefficient, [m$^{-1}$]
$\rho_F$    – density of fuel, [kgm$^{-3}$]
$\rho_w$    – density of water, [kgm$^{-3}$]
$\rho_\infty$    – density of fire environments (*i. e.* the surrounding air), [kgm$^{-3}$]
$\tau_0$    – hypothetic time for complete burning of the fuel layer without boilover, ($\tau_0 = y_0/V_a$), [s]
$\Phi$    – flux released at the surface (eq. 16) due to radiation and phase change, [Wm$^{-2}$s]
    – radiative fraction of the radiation feedback to the pool surface), [–]

*Dimensionless groups*

Bu                       – Bouger number, [–]
$B_{SA} = \dot{q}_s y_0/\lambda_F(T_s - T_\infty)$   – dimensionless number
Fo $= y_0 t/a_F$           – Fourier number, [–]
Fo$^e = y_0 t_{b0}/a_F$       – Fourier number at the boilover onset
$H_p = \dot{q} / \dot{m} H_V$        – dimensionless number
$N_0 = \dot{q}_s y_0/\lambda_F(T_s - T_\infty)$   – dimensionless group (radiation-conduction number)
$N_{DHS} = y_0 v(t)/a_F = y_0 V_a/a_F$   – dimensionless group (pseudo Peclet number)
Ste $= C_{pF}(T_s - T_\infty)/H_V$    – Stefan number, [–]

*Subscripts*                          *Superscripts*

c    – conductivity            f    – flame
F    – fuel
f    – flame                    *Special symbols*
s    – surface
$v$    – vapour                 – proportional to
w    – water                  – order of magnitude
W–SH   – water superheat      ~    – about equal
$\infty$    – ambient conditions

Author's address:

*J. Hristov*

Department of Chemical Engineering,
University of Chemical Technology and Metallurgy
8, Kl. Ohridsky blvd., 1756 Sofia, Bulgaria

E-mail: jyh@uctm.edu, jordan.hristov@mail.bg., hristovmeister@gmail.com